# Anomalous wave transport in one-dimensional systems with Lévy disorder

Xujun Ma [1,2] and Azriel Z. Genack [1,2] *

[1]Department of Physics, Queens College of the City University of New York, Flushing, NY, 11367 USA

[2]Physics Program, Graduate Center of the City University of New York, New York, NY, 10016 USA

* Corresponding author. Email: Azriel.Genack@qc.cuny.edu

**We use random matrix theory to study the statistics of wave transport in one-dimensional random media with Lévy disorder, in which waves perform Lévy flights. We discover that the mean transmission scales asymptotically with system length $L$ as a power law, $\langle T \rangle \to L^{-\alpha}$, where $\alpha$ is the stability index of the Lévy distribution, and $\langle \ln T \rangle$ scales as a power law, $\langle \ln T \rangle \propto -L^\alpha$. We find the average logarithm of intensity falls off as a power law inside the system, $\langle \ln I(x) \rangle \propto -x^\alpha$, and obtain the analytical form of average intensity $\langle I(x) \rangle$ at any depth $x$.**

Lévy flights are a class of random processes that occur widely and are of fundamental importance in physics [1-3], especially for anomalous diffusion and fractional kinetics. They also play an important role in other disciplines such as economics [4], biology [5] and seismology [6]. The step lengths of Lévy flights follow the α-stable distribution (or Lévy distribution), which is heavy-tailed with a probability density function (PDF) that decays as a power law for large $X$, $p(X) \propto \frac{1}{X^{1+\alpha}}$ with $0 < \alpha < 2$. As a result, the second moment of $p(X)$ diverges for $0 < \alpha < 2$ as does the first moment for $0 < \alpha < 1$. It is a special case for $\alpha=2$ that $p(X)$ is a Gaussian function and the random walks follow Brownian motion statistics [7].

Recently, the Lévy flights for light were experimentally investigated. Barthelemy *et al.* [8] created an optical material that consists of a random packing of glass microspheres with a Lévy distribution of diameters and scattering particles of titanium dioxide filled in between, which they call a *Lévy glass*. Light undergoes ballistic propagation inside glass microspheres and scatters at the spherical boundaries. Therefore, the light trajectories are Lévy flights. This work opened a window to the study of Lévy flights for coherent waves and motivated theoretical interests on wave transport in random media with Lévy disorder [9-15].

Anderson localization plays a central role in the study of wave transport in random media. It was originally introduced in quantum electron transport that as a consequence of destructive interference in multiple scattering, the electron wave function decays exponentially inside the random system. Anderson localization is fundamentally a wave effect that applies to both classical waves and quantum waves. Random matrix theory has been successfully applied to study the statistical properties of wave transport quantities that characterize Anderson localization effect [16]. The Dorokhov-Mello-Pereyra-Kumar (DMPK) equation is a partial differential equation that the PDF of transmission satisfies. Its solution gives the PDF of transmission, from which the average of transmission is found, $\langle T \rangle \propto \exp\left(-\frac{L}{2l}\right)$, and also the average of the logarithm of transmission $\langle \ln T \rangle = -\frac{L}{l}$, where $L$ is system length and $l$ is the mean free path. Random matrix theory has also been applied to study the statistical properties of intensity in the interior of random systems. The average of the logarithm of intensity inside a system falls linearly with depth $x$, $\langle \ln I(x) \rangle = -\frac{x}{l}$ [17], and the analytical expression of mean intensity profile $\langle I(x) \rangle$ was obtained in Ref. [18].

However, the nature of wave localization might have an entirely different form when waves perform Lévy flights in random media. In this Letter, we show that anomalous Anderson localization occurs in 1D systems with Lévy disorder, in which the wave trajectories are Lévy flights. We use random matrix theory to study the statistical properties of transmission and find the average logarithm of transmission scales as a power law with system length, $\langle \ln T \rangle \propto -L^\alpha$, and the mean transmission scales as a power law for large $L$, $\langle T \rangle \to L^{-\alpha}$. We also investigate the statistics of intensity inside the system and obtain the analytical form of $\langle I(x) \rangle$, and find the average logarithm of intensity falls as a power law with depth $x$, $\langle \ln I(x) \rangle \propto -x^\alpha$. These statistical properties are different from standard Anderson localization described above and reveal the anomalous transport signatures.

A 1D random system with Lévy disorder is shown in Fig. 1. The spacings between two nearest neighbor scatterers follow a Lévy distribution. Waves perform ballistic transport between scatterers and undergo multiple scattering before leaving the system, with the trajectories being Lévy flights. We did computer simulations to mimic electromagnetic wave transport in this system. In our simulations, a plane electromagnetic wave impinges normally on an statistically equivalent ensemble of random layered structures with alternating indices of refraction $n_{A(B)}$. The thicknesses of the layers are random variables follow Lévy distribution. Waves perform free propagation inside layers with Lévy distributed step lengths and scatter at the interfaces of layers due to the mismatch of indices of refraction. We set $n_A = 1$ and draw $n_B$ from a uniform distribution to obtain random scattering strengths. The intensity inside the sample and the transmission at output are computed via scattering matrix.

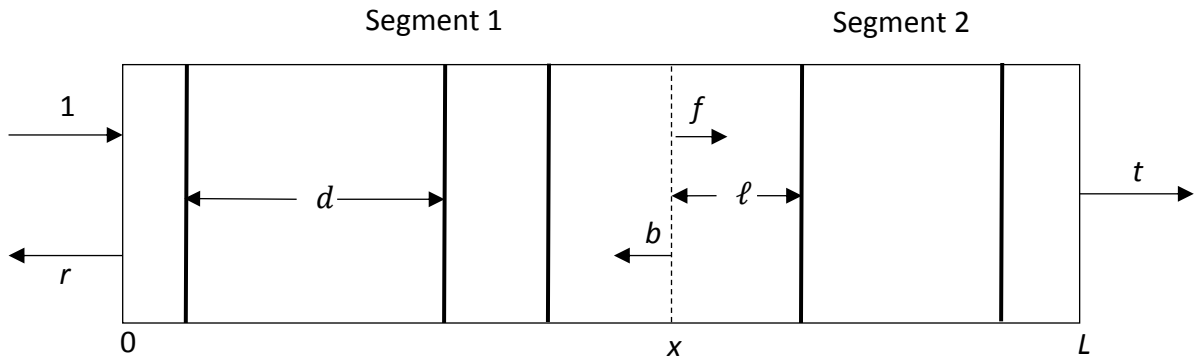

Fig. 1 Schematic of a 1D random system with Lévy disorder. The spacing of scatterers $d$ follows a Lévy distribution. The amplitude of incident wave is 1, and $r$ and $t$ are the amplitudes of reflected and transmitted waves. In order to study the intensity inside the system at depth $x$, we separate the system at $x$ into two segments. Waves moving forward and backward at $x$ are denoted by $f$ and $b$. The distance between $x$ and the first scatterer in segment 2 is $\ell$.

There is no closed form expression for the PDF of α-stable distribution, except for a few special cases. Nevertheless, $p(X)$ can be obtained by taking the inverse Fourier transform of their characteristic function [7],

$$E(\exp(i\theta X)) = \begin{cases} \exp\left\{-\sigma^\alpha |\theta|^\alpha \left(1 - i\beta(\text{sign }\theta)\tan\frac{\pi\alpha}{2}\right) + i\mu\theta\right\} & \text{if } \alpha \neq 1, \\ \exp\left\{-\sigma|\theta|\left(1 + i\beta\frac{2}{\pi}(\text{sign }\theta)\ln|\theta|\right) + i\mu\theta\right\} & \text{if } \alpha = 1. \end{cases} \quad (1)$$

Here $\alpha \in (0, 2]$, $\beta \in [-1, 1]$, $\sigma \in (0, \infty)$ and $\mu \in (-\infty, \infty)$ are the stability parameter, skewness parameter, scale parameter and location parameter respectively. Since the spacings between scatterers are positive, we work on one-sided Lévy distribution with $0 < \alpha < 1$, $\beta=1$, $0 < \sigma < \infty$ and $\mu=0$. Because the randomness in our systems is stationary or quenched, waves travel the same step length after a backscattering, while Lévy flights have independent subsequent step lengths ("annealed" disorder). The Lévy flights exponent $\alpha'$ in quenched case is related to the annealed value $\alpha$ by $\alpha' = \alpha + \left(\frac{2}{m}\right)\max(0, \alpha - m)$ in $m$-dimension. Thus, for the one-sided Lévy flights in 1D, we have $\alpha'= \alpha$ [19]. Note that Lévy flights with finite velocity are called Lévy walks. There is no particular difference between a Lévy flight and a Lévy walk in this Letter because all physical quantities and system configurations are time independent.

We first introduce the statistics of the number of scatterers $n$ in a system of length $L$. It is actually equivalent to the question of how many steps it takes for a one-sided Lévy flight starting at 0 to pass position $L$. This is exactly the first passage time problem of Lévy flights that is well studied by Klafter *et al.* [3, 20-21]. Denote the PDF of $n$ in a system with length $L$ as $P_L(n)$ and the expectation value as $E_L(n)$. According to Ref. [3, Chapter (7)], we have

$$P_L(n) = \frac{L}{\alpha \sigma n^{1+\frac{1}{\alpha}}} q_{\alpha,1}\left(\frac{L}{\sigma n^{\frac{1}{\alpha}}}\right), \quad (2)$$

where $q_{\alpha,\sigma}(X)$ is the PDF of a one-sided Lévy distribution with exponent $\alpha$ and scale parameter $\sigma$. Its expectation value is given by

$$E_L(n) = \frac{\cos(\pi\alpha/2)}{\Gamma(1+\alpha)\sigma^\alpha} L^\alpha, \qquad (3)$$

where $\Gamma$ is the Gamma function.

From the standard scaling theory of localization, for a random media with fixed number of scatterers $n$, the average of the logarithm of transmission is proportional to $n$: $\langle -\ln T \rangle_n = an$, where $a$ is a positive constant and is related to the average scattering strength [22]. By the law of total expectation, we obtain

$$\langle \ln T \rangle = E(\langle \ln T \rangle_n | n) \qquad (4a)$$

$$= -aE_L(n) \qquad (4b)$$

$$= -\frac{a \cos\left(\frac{\pi\alpha}{2}\right)}{\Gamma(1+\alpha)\sigma^\alpha} L^\alpha. \qquad (4c)$$

This analytical result is verified in simulation plotted in Fig. 2. This indicates that $\langle \ln T \rangle$ scales with the length of the system as $L^\alpha$ instead of linearly with $L$ as in standard Anderson localization.

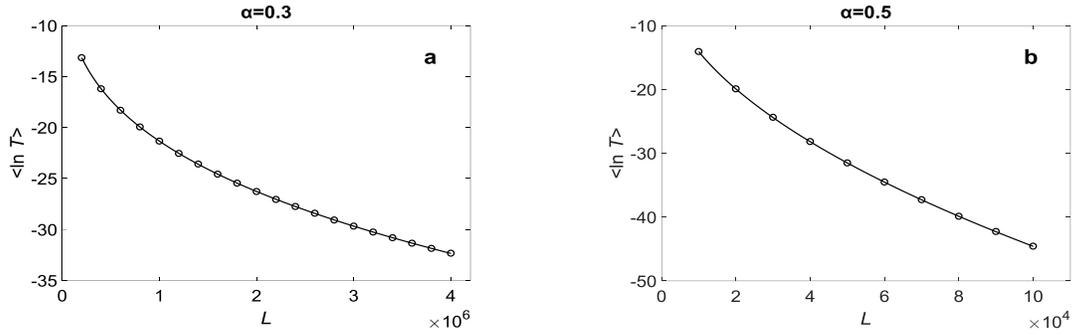

Fig. 2 The ensemble average of $\ln T$ scales as a power law with exponent $\alpha$ for (**a**) $\alpha=0.3$. (**b**) $\alpha=0.5$. The analytical result of Eq. (4c) (solid line) match simulation data (circles).

With the analytical form of $\langle \ln T \rangle$, we can calculate $\langle \ln I(x) \rangle$ at any depth $x$ inside this system. We separate the system into two segments shown in Fig. 1. Denote the scattering matrices and transmissions of these two segments as $S_1$, $S_2$ and $T_1$, $T_2$ respectively. We show in Supplementary Materials that

$$\langle \ln I(x) \rangle = \langle \ln T \rangle - \langle \ln T_2 \rangle. \tag{5}$$

The first term at the left side is given by Eq. (4c) and the derivation of the second term involves the first passage leapover of Lévy flights. As described above, we regard the construction of the scatters in segment 1 as the first passage time problem of a Lévy flight starting at 0 and passing $x$. The last flight does not reach position $x$ exactly. Instead, it leaps over $x$ by a distance $\ell$, as shown in Fig. 1. For this reason, the effective length of segment 2 in which all scatterers are contained is $L - x - \ell$. The PDF of $\ell$ for a one-sided Lévy flight with target at $x$ is (Ref. [20, Eq. (26)])

$$P(\ell) = \frac{\sin(\pi\alpha)}{\pi} \frac{x^\alpha}{\ell^\alpha(x+\ell)}. \tag{6}$$

Now, by Eq. (4b), we have

$$\langle \ln T_2 \rangle = -aE_{L-x-\ell}(n) \tag{7a}$$

$$= -aE(E_{L-x-\ell}(n|\ell)|\ell) \tag{7b}$$

$$= -a \int_0^{L-x} \frac{\cos\left(\frac{\pi\alpha}{2}\right)}{\Gamma(1+\alpha)\sigma^\alpha} (L-x-\ell)^\alpha \frac{\sin(\pi\alpha)}{\pi} \frac{x^\alpha}{\ell^\alpha(x+\ell)} d\ell \tag{7c}$$

$$= -\frac{a \cos\left(\frac{\pi\alpha}{2}\right)}{\Gamma(1+\alpha)\sigma^\alpha} (L^\alpha - x^\alpha). \tag{7d}$$

We used Eq. (3.228) of Ref. [24] to evaluate the integral in Eq. (7c). Note that the upper limit of this integral is $L - x$, because for $\ell > L - x$, there are no scatterers inside segment 2, thus $T_2 = 1$, $\ln T_2 = 0$. Substituting Eq. (4c) and Eq. (7d) into Eq. (5) gives

$$\langle \ln I(x) \rangle = -\frac{a \cos\left(\frac{\pi\alpha}{2}\right)}{\Gamma(1+\alpha)\sigma^\alpha} x^\alpha. \tag{8}$$

This analytical result is in agreement with simulation shown in Fig. 3. This shows the anomalous transport behavior of waves in Lévy disorder systems. The ensemble average of $\ln I(x)$ decays as a power law with exponent α in contrast to linear decay in standard Anderson localization. As $x \to L$, $\langle \ln I(x) \rangle \to -\frac{a \cos\left(\frac{\pi\alpha}{2}\right)}{\Gamma(1+\alpha)\sigma^\alpha} L^\alpha$, which is in agreement with Eq. (4c).

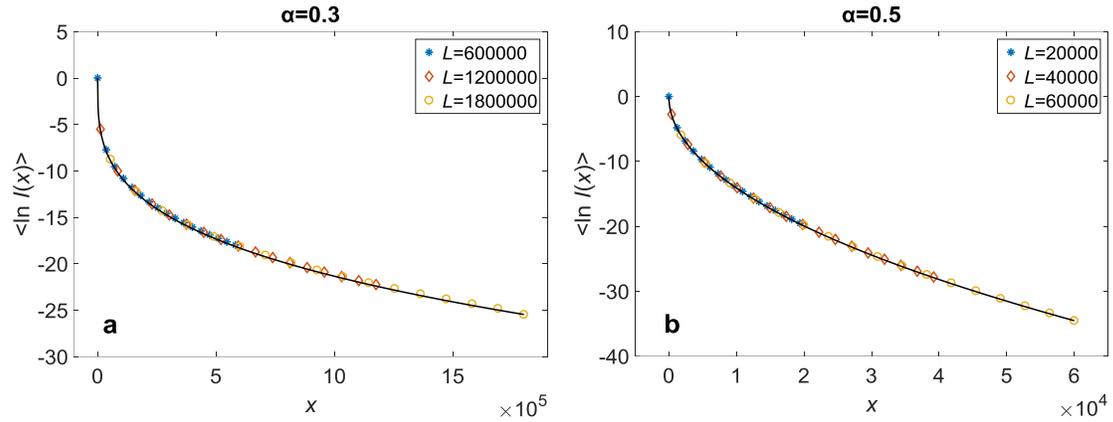

Fig.3 $\langle \ln I(x) \rangle$ falls off as a power law with exponent $\alpha$ for (**a**) $\alpha$=0.3. (**b**) $\alpha$=0.5. The solid lines obtained from Eq. (8) are in agreement with simulation data (symbols) for different sample lengths.

Before deriving the expression for the mean transmission $\langle T \rangle$ of this system, we introduce the PDF of transmission for wave transport in random media. The DMPK equation, which is derived from random matrix theory by applying a maximum entropy approach, is a Fokker-Plank equation of the PDF of transmission $P_s(\lambda)$ [23]. In 1D, the DMPK equation reduces to Melnikov's equation,

$$\frac{\partial P_s(\lambda)}{\partial s} = \frac{\partial}{\partial \lambda}\left[\lambda(\lambda+1)\frac{\partial P_s(\lambda)}{\partial \lambda}\right], \tag{9}$$

where $\lambda = \frac{1-T}{T}$ and $s = -\frac{L}{l} = \langle -\ln T \rangle = an$. Note that the relation $s = -\frac{L}{l}$ is defined for standard Anderson localization, in which the PDF of spacings between scatterers has finite first moment, thus the average density of scatterers is a constant, and so is $l$. For a random system with Lévy disorder, the average density of scatterers changes inside the system, so $l$ is not defined, or it is not a constant. However, the relation $s = an$ holds for both cases [9]. The solution of Eq. (9) can be written as

$$P_s(T) = \frac{s^{-\frac{3}{2}} e^{-\frac{s}{4}}}{\sqrt{2\pi}\, T^2} \int_{y_0}^{\infty} dy \frac{y e^{-\frac{y^2}{4s}}}{\sqrt{\cosh y + 1 - 2/T}}, \quad (10)$$

where $y_0 = \text{arcosh}\left(\frac{2}{T}\right) - 1$. We point out that the distribution of transmission is determined by a single parameter $s = an$, where $a$ is a constant. Now we can calculate the average transmission for a 1D Lévy disordered system with a given number of scatterers $n$:

$$\langle T_n \rangle = \int_0^1 T P_{an}(T)\, dT = \int_0^1 dT \frac{(an)^{-\frac{3}{2}} e^{-\frac{an}{4}}}{\sqrt{2\pi}\, T} \int_{y_0}^{\infty} dy \frac{y e^{-\frac{y^2}{4an}}}{\sqrt{\cosh y + 1 - 2/T}}. \quad (11)$$

and the average transmission $\langle T \rangle$ can be derived by integrating $\langle T_n \rangle$ over $n$ multiplying the distribution of $n$ given by Eq. (2),

$$\langle T \rangle = \int_0^{\infty} P_L(n) \langle T_n \rangle dn$$

$$= \int_0^{\infty} dn \frac{L}{a\sigma n^{1+\frac{1}{\alpha}}} q_{\alpha,1}\left(\frac{L}{\sigma n^{\frac{1}{\alpha}}}\right) \int_0^1 dT \frac{(an)^{-\frac{3}{2}} e^{-\frac{an}{4}}}{\sqrt{2\pi}\, T} \int_{y_0}^{\infty} dy \frac{y e^{-\frac{y^2}{4an}}}{\sqrt{\cosh y + 1 - 2/T}}. \quad (12)$$

The scaling of the mean transmission for 1D Lévy disordered systems with different values of $\alpha$ is plotted in Fig. 4. The solid lines representing the analytical results are evaluated by carrying out the integrals in Eq. (12) numerically. These results are in agreement with simulations. It is known from the scaling theory of localization that the average transmission decays exponentially for deeply localized systems, $\langle T_s \rangle \propto e^{-\frac{L}{2l}} = e^{-s/2}$. As a result of the heavy-tail property of $q_{\alpha,1}(X)$, $q_{\alpha,1}\left(\frac{L}{\sigma n^{\frac{1}{\alpha}}}\right) \propto \left(\frac{L}{\sigma n^{\frac{1}{\alpha}}}\right)^{-(1+\alpha)}$, then $P_L(n) = \frac{L}{\alpha \sigma n^{1+\frac{1}{\alpha}}} q_{\alpha,1}\left(\frac{L}{\sigma n^{\frac{1}{\alpha}}}\right) \propto \frac{1}{\alpha}\left(\frac{L}{\sigma}\right)^{-\alpha}$. Thus, for large $L$,

$$\langle T \rangle \sim \int_0^\infty \frac{1}{\alpha}\left(\frac{L}{\sigma}\right)^{-\alpha} e^{-an/2} dn = \frac{2}{\alpha a}\left(\frac{L}{\sigma}\right)^{-\alpha} \to L^{-\alpha}, \tag{13}$$

the mean transmission scales as a power law. This is in agreement with simulation results as shown in the log-scale plot of $\langle T \rangle$ vs. $L$ in Fig. 4. We observe that, for large $L$, the curves tend to fall linearly with slope $-\alpha$.

The average of the intensity inside the system can also be derived using random matrix theory. We first give a general expression for $\langle I(x) \rangle$ for waves inside 1D random media and then apply it to random systems with Lévy disorder. The expression of $\langle I(x) \rangle$ is (see Supplementary Materials)

$$\langle I(x) \rangle = \int_0^1 \int_0^1 \frac{2T_1 - T_1 T_2}{T_1 + T_2 - T_1 T_2} P_{S_1}(T_1) P_{S_2}(T_2) dT_1 dT_2. \tag{14}$$

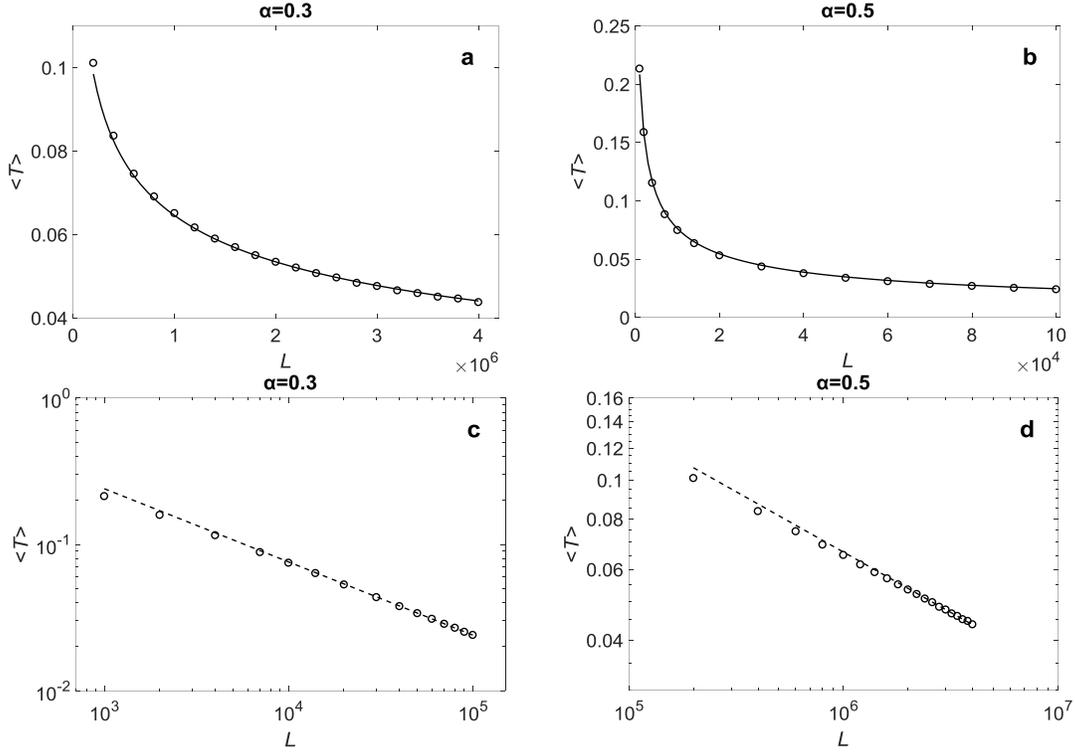

Fig. 4 (**a**) and (**b**) The solid lines represent the analytical result of $\langle T \rangle$, given by Eq. (12), match simulation data (circles) for $\alpha=0.3$ and 0.5. (**c**) and (**d**) Simulation data (circles) of $\langle T \rangle$ vs. $L$ are plotted in log-scale. Dashed straight lines with slope 0.3 and 0.5 match data points for large $L$. It indicates that $\langle T \rangle$ scales with system length as a power law with exponent $-\alpha$ asysmtotically.

We point out that this expression applies to both standard and anomalous transport of waves in 1D random media. For standard random media, the parameter $s$ is proportional to length with slope the inverse of mean free path, thus $s_1 = \frac{x}{l}$, $s_2 = \frac{L-x}{l}$. For random systems with Lévy disorder, the proportionality does not hold but we can still use $s_1 = \langle -\ln T_1 \rangle = an_1$, $s_2 = \langle -\ln T_2 \rangle = an_2$. Here $n_1$ and $n_2$ are random variables with their PDF given by Eq. (2). We also need to consider the first passage leapover of Lévy flights as described above (shown in Fig. 1.).

If the first passage leapover $\ell$ is less than the length of segment 2, $\ell < L - x$, scatterers exist in segment 2. However, if $\ell > L - x$, segment 2 is empty, thus $T_2 = 1$, $\frac{2T_1 - T_1 T_2}{T_1 + T_2 - T_1 T_2} = T_1$. The probabilities of these two cases can be obtained by integrating Eq. (6) with respect to $\ell$ from 0 to $L - x$ and $L - x$ to infinity respectively. We can express $\langle I(x) \rangle$ as:

$\langle I(x) \rangle$

$$= \int_0^{L-x} \frac{\sin(\pi\alpha)}{\pi} \frac{x^\alpha}{\ell^\alpha (x+\ell)} d\ell \int_0^\infty dn_1 \int_1^\infty dn_2 \int_0^1 dT_1 \int_0^1 dT_2 \frac{2T_1 - T_1 T_2}{T_1 + T_2 - T_1 T_2} P_{an_1}(T_1) P_{an_2}(T_2) P$$

$$+ \int_{L-x}^\infty \frac{\sin(\pi\alpha)}{\pi} \frac{x^\alpha}{\ell^\alpha (x+\ell)} d\ell \times \int_0^\infty dn_1 \int_0^1 dT_1 \, T_1 P_{an_1}(T_1) P_x(n_1). \tag{15}$$

The first term represents the case $\ell < L - x$ where segment 2 contains at least one scatterers so the lower limit of the integral with respect to $n_2$ is 1. $P_{an_1}(T_1)$ and $P_{an_2}(T_2)$ are given by Eq. (10). $P_x(n_1)$ and $P_{L-x-\ell}(n_2)$ are the probabilities of the numbers of scatterers in each segment given by Eq. (2). Again, the effective length of segment 2 is $L - x - \ell$. The second term stands for the case $\ell > L - x$ in which segment 2 is empty, and the second integral is just the mean transmission of segment 1, which can also be obtained by replacing $L$ with $x$ in Eq. (12). We have not been able to evaluate the integral of the first term of Eq. (15) numerically and therefore present the simulation results in Fig. 5.

We check Eq. (15) at the boundaries of a system. At the input edge, $x = 0$, $n_1 = 0$, $T_1 = 1$, the function $\frac{x^\alpha}{\ell^\alpha (x+\ell)}$ is nonzero only for $\ell \to 0$ so $\int_0^{L-x} \frac{x^\alpha}{\ell^\alpha (x+\ell)} d\ell \to 1$ and $\int_{L-x}^\infty \frac{x^\alpha}{\ell^\alpha (x+\ell)} d\ell \to 0$.

$P_{L-x-\ell}(n_2) \to P_L(n_2)$ and segment 2 is the whole system thus $T_2$ is equivalent to $T$. Then Eq. (15) goes to $\langle I(0) \rangle = \int_0^\infty dn_2 \int_0^1 dT_2 (2-T_2) P_{an_2}(T_2) P_L(n_2) = 2 - \langle T \rangle = 1 + \langle R \rangle$. Here $\langle T \rangle$ and $\langle R \rangle$ are the average transmission and reflection of the system. The last equation is required by flux conservation. Indeed, $\langle I(0) \rangle = \langle |1+r|^2 \rangle = 1 + \langle R \rangle$ [18]. At the output, the first term is zero and the first integral of the second term is 1. Now segment 1 is the whole system, so $T_1$ is equivalent to $T$. Thus $\langle I(L) \rangle = \langle T_1 \rangle = \langle T \rangle$.

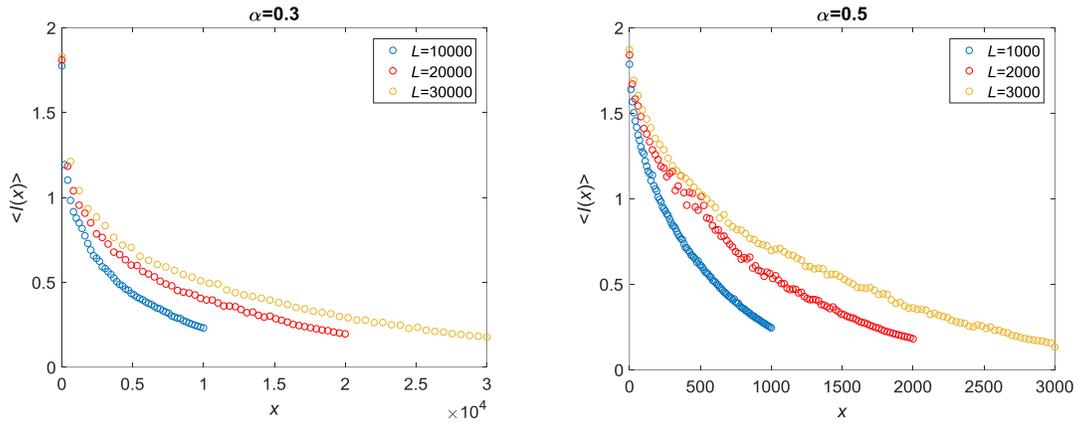

Fig. 5 Simulation results of $\langle I(x) \rangle$ for wave transport in 1D Lévy disordered systems of different lengths with $\alpha$=0.3 and 0.5.

**Summary**

Lévy flights occur widely in nature. The Lévy flights of classical particles have been intensively investigated while the study of Lévy flights of coherent waves has emerged only recently. We have used random matrix theory to study the statistics of wave transport in 1D random systems with Lévy disorder, in which waves perform Lévy flights. We obtain the scaling of transport quantities. Also, we study the intensity profile inside the system for the first time. We hope these results in 1D would be the basis for future investigation of higher-dimensional Lévy flights of waves. Moreover, we find that anomalous Anderson localization occurs in our system. This gives us a better understanding of the effect of disorder in Anderson localization and might stimulate researches on customizing transmission and intensity distribution by controlling the disorder.

**Acknowledgement**

The authors acknowledge the support by the National Science Foundation (NSF/DMR/-BSF: 1609218).

**Supplementary Materials**

**Proof of Eq. (5)**

We express the scattering matrices of the two segments as

$$S_i = \begin{pmatrix} r_i & t'_i \\ t_i & r'_i \end{pmatrix}, \quad i = 1, 2. \tag{S.1}$$

From the definition of scattering matrix, we have

$$S_2 \begin{pmatrix} f \\ 0 \end{pmatrix} = \begin{pmatrix} b \\ t \end{pmatrix}, \tag{S.2}$$

which leads to

$$f = \frac{t}{t_2}, \quad b = \frac{r_2}{t_2} t. \tag{S.3}$$

The intensity at $x$ is

$$I(x) = |f + b|^2 = |\frac{t}{t_2} + \frac{r_2}{t_2} t|^2 = T|\frac{1}{t_2} + \frac{r_2}{t_2}|^2, \tag{S.4}$$

thus

$$\ln I(x) = \ln T + \ln |\frac{1}{t_2} + \frac{r_2}{t_2}|^2. \tag{S. 5}$$

These scattering matrices can be expressed in polar representation under time-reversal invariance symmetry as [23]:

$$S_j = \begin{pmatrix} r_j & t'_j \\ t_j & r'_j \end{pmatrix} = \begin{pmatrix} e^{i\phi_j} & 0 \\ 0 & e^{i\psi_j} \end{pmatrix} \begin{pmatrix} -\sqrt{1-T_j} & \sqrt{T_j} \\ \sqrt{T_j} & \sqrt{1-T_j} \end{pmatrix} \begin{pmatrix} e^{i\phi_j} & 0 \\ 0 & e^{i\psi_j} \end{pmatrix}$$

$$= \begin{pmatrix} -\sqrt{1-T_j}e^{i2\phi_j} & \sqrt{T_j}e^{i(\phi_j+\psi_j)} \\ \sqrt{T_j}e^{i(\phi_j+\psi_j)} & \sqrt{1-T_j}e^{i2\psi_j} \end{pmatrix}, \quad j=1,2. \tag{S. 6}$$

Here $T_j$ are the transmissions of these two segments, and $\phi_j$ and $\psi_j$ are random phases. This gives

$$\ln |\frac{1}{t_2} + \frac{r_2}{t_2}|^2 = \ln \frac{1}{T_2}(2 - T_2 - 2\sqrt{1-T_2}\cos(2\phi_2)). \tag{S. 7}$$

Substituting this result into Eq. (S. 5) and taking the average of both sides with the identity (see Eq. (4.224.9) of [24]):

$$\int_0^\pi d\theta \ln(A + B\cos\theta) = \pi \ln\frac{1}{2}\left[A + (A^2 - B^2)^{\frac{1}{2}}\right], \tag{S. 8}$$

which equals $\frac{1}{T_2}$ for our $A = \frac{2-T_2}{T_2}$, $B = -\frac{2\sqrt{1-T_2}}{T_2}$, we have

$$\langle \ln I(x) \rangle = \langle \ln T \rangle + \int_0^1 \int_0^{2\pi} \frac{\mathrm{d}\phi_2}{2\pi} \ln(A + B\cos\phi_2) P(T_2) \mathrm{d}T_2$$

$$= \langle \ln T \rangle + \int_0^1 \ln \frac{1}{T_2} P(T_2) \mathrm{d}T_2$$

$$= \langle \ln T \rangle - \langle \ln T_2 \rangle. \tag{S. 9}$$

**Proof of Eq. (14)**

We start from

$$I(x) = |f + b|^2 = \left|\frac{t}{t_2} + \frac{r_2}{t_2}t\right|^2 = T\left|\frac{1}{t_2} + \frac{r_2}{t_2}\right|^2. \tag{S. 10}$$

Using the polar representation of $S_2$ according to Eq. (S. 6), we have

$$\left|\frac{1}{t_2} + \frac{r_2}{t_2}\right|^2 = \frac{1}{T_2}\left(2 - T_2 - 2\sqrt{1-T_2}\cos(2\phi_2)\right). \tag{S. 11}$$

The law of composition of $S_1$ and $S_2$ gives:

$$t = \frac{t_1 t_2}{1 - r_1' r_2} = \frac{\sqrt{T_1 T_2} e^{i(\phi_1 + \psi_1 + \phi_2 + \psi_2)}}{1 + \sqrt{(1-T_1)(1-T_2)} e^{i(\psi_1 + \phi_2)}}. \tag{S. 12}$$

Thus

$$T = tt^* = \frac{T_1 T_2}{2 - T_1 - T_2 + T_1 T_2 + 2\sqrt{(1-T_1)(1-T_2)}\cos(2(\psi_1 + \phi_2))}. \quad \text{(S. 13)}$$

Substituting Eq. (S. 10) and Eq. (S. 13) into Eq. (S. 4) gives

$$I(x) = \frac{T_1(2 - T_2 - 2\sqrt{1-T_2}\cos(2\phi_2))}{2 - T_1 - T_2 + T_1 T_2 + 2\sqrt{(1-T_1)(1-T_2)}\cos(2(\psi_1 + \phi_2))}. \quad \text{(S. 14)}$$

We make the change of variables $\mu = \psi_1 + \phi_2$ and take the average of both sides of Eq. (S. 14):

$$\langle I(x) \rangle = \int_0^1 \int_0^1 dT_1 dT_2 P_{s_1}(T_1) P_{s_2}(T_2) \bar{I}(T_1, T_2, \phi_2, \mu), \quad \text{(S. 15)}$$

where

$$\bar{I}(T_1, T_2, \phi_2, \mu)$$
$$= \int_0^{2\pi} \int_0^{2\pi} d\phi_2 d\mu \frac{T_1(2 - T_2 - 2\sqrt{1-T_2}\cos(2\phi_2))}{2 - T_1 - T_2 + T_1 T_2 + 2\sqrt{(1-T_1)(1-T_2)}\cos(2\mu)} \quad \text{(S. 16)}$$
$$= \frac{2T_1 - T_1 T_2}{T_1 + T_2 - T_1 T_2}.$$

In Eq. (S. 16), the integral with respect to $\phi_2$ vanishes, and the integral of $\mu$ is carried out using Eq. (3613.1) of Ref. [24]. Thus, we get

$$\langle I(x) \rangle = \int_0^1 \int_0^1 \frac{2T_1 - T_1 T_2}{T_1 + T_2 - T_1 T_2} P_{s_1}(T_1) P_{s_2}(T_2) dT_1 dT_2. \quad \text{(S. 17)}$$